\documentclass[aps,prd,amssymb,nobibnotes, amsmath,nofootinbib,superscriptaddress,floatfix,preprintnumbers]{revtex4-1}

\usepackage{graphicx,color}
\usepackage{dcolumn}
\usepackage{bm}
\usepackage{hyperref}
\usepackage{amsmath}
\usepackage{mathtools}
\usepackage{multirow}
\usepackage{stackrel}
\usepackage[lofdepth,lotdepth,caption=false]{subfig}
\usepackage{tikz-feynman}
\usepackage{verbatim}
\usepackage{xcolor}
\def\d{{\rm d}}

\begin{document}


\begin{flushleft}
LAPTH-014/24
\end{flushleft}

\title{A closer scrutiny of cosmic ray proton energy losses}

\author{AmirFarzan Esmaeili}
\email{a.farzan.1993@aluno.puc-rio.br}
\affiliation{Departamento de Física, Pontifícia Universidade Católica do Rio de Janeiro, Rio de Janeiro 22452-970, Brazil}

\author{Arman Esmaili}
\email{arman@puc-rio.br}
\affiliation{Departamento de Física, Pontifícia Universidade Católica do Rio de Janeiro, Rio de Janeiro 22452-970, Brazil}

\author{Pasquale Dario~Serpico}
\email{serpico@lapth.cnrs.fr}
\affiliation{LAPTh, CNRS, USMB, F-74940 Annecy, France}



\begin{abstract}
The percent-level precision attained by modern cosmic ray (CR) observations motivates reaching a comparable or better control of theoretical uncertainties. Here we focus on energy-loss processes affecting low-energy CR protons ($\sim 0.1-5$ GeV), where the experimental errors are small and collisional effects play a comparatively larger role with respect to collisionless transport ones. We study three aspects of the problem: i) We quantitatively assess the role of the nuclear elastic cross-section, for the first time, providing  analytical formulae for the stopping power and inelasticity. ii) We discuss the error arising from treating both elastic and pion production inelastic interactions as continuous energy loss processes, as opposed to catastrophic ones. The former is the approximation used in virtually all modern numerical calculations. iii) We consider sub-leading effects such as relativistic corrections, radiative and medium processes in ionization energy-losses. Our analysis reveals that neglecting i) leads to errors close to 1\%, notably around and below 1 GeV; neglecting ii) leads to errors reaching about 3\% within the considered energy range; iii) contributes to a minor effect, gauged at the level of 0.1\%. Consequently, while iii) can currently be neglected, ii) warrants consideration, and we also recommend incorporating i) into computations. We conclude with some perspectives on further steps to be taken towards a high-precision goal of theoretical CR predictions regarding the treatment of energy-losses.                                  
\end{abstract}
\maketitle

\section{Introduction}\label{sec:intro}
Over the past decade, a wealth of balloon-borne and space-borne detectors has ushered the study of Galactic cosmic rays (CR) into a precision era, stimulating numerous new questions on the underlying physics (for reviews, see e.g.~\cite{Serpico:2015caa,Serpico:2018lkb,Boezio:2020pyu,Moskalenko:2023mjl}). This also allows one to tackle  some important applications in astroparticle physics with a sharpened diagnostic tool: A notable example is provided by the impact that a refined understanding and error budget of antiproton CR~\cite{Winkler:2017xor,Boudaud:2019efq,Cuoco:2019kuu,Heisig:2020nse} has on constraints on weakly interacting dark matter particles~\cite{DiMauro:2021qcf,Calore:2022stf,Lv:2023gdt,DelaTorreLuque:2024ozf}.

 In particular, we have recently witnessed both a remarkable reduction of errors and an extension of the dynamical range covered by the data: In these respects, AMS-02~\footnote{\texttt https://ams02.space} measured proton fluxes have a statistical precision often below the permille level, and the overall systematic errors are typically estimated at the 1\% level~\cite{AMS:2015tnn}. Additionally, the Voyager mission has provided a notable advance by acquiring sub-GeV CR data  beyond the heliosphere for the first time, free of solar modulation effects~\cite{2013Sci...341..150S}. 
To get the most out of the newly acquired diagnostic power, this reduction in the size of experimental errors  must be accompanied by a refinement in theoretical errors. 

One important ingredient in that direction consists in nuclear, hadronic and particle physics cross sections (for the context of collisional vs. collisionless effects shaping the CR dynamics, see e.g. the lecture notes~\cite{Serpico:2023yag}). The limitations of current libraries, ultimately due to scarce and sparse data, are well-known, and extensively studied for spallation cross-sections~\cite{Moskalenko:2004fe,Genolini:2018ekk,Genolini:2023kcj}. This has stimulated meetings between collider physicists, CR experimentalists, and theorists to advance knowledge in this direction, such as the \texttt{Cross sections for Cosmic Rays} series at CERN~\footnote{The website of the latest in the series is at https://indico.cern.ch/event/1377509/}.
In this context, a more overlooked direction consists in assessing how much of {\it known} physics is actually included in current CR treatments. In the past decade, some efforts have been made towards a better modeling of yields of secondary particles in CR collisions, see e.g.~\cite{Mazziotta:2015uba,Orusa:2023bnl,DiMauro:2023oqx}. 
Here we tackle a somewhat complementary avenue towards this precision goal, focusing on the description of {\it energy-losses}, along three directions:

i) We revisit the processes affecting protons in the low-energy regime in particular below 5 GeV, which is where losses are  most relevant compared to diffusive transport phenomena, and CR data have the best precision. We provide for the first time a realistic assessment of proton-proton nuclear elastic energy losses, only qualitatively gauged as unimportant in the past (see e.g. the comments in \cite{Bergstrom:1999jc} or on the \href{https://dmaurin.gitlab.io/USINE/input_xs_data.html?highlight=elastic}{\texttt{USINE} propagation code webpage}).  We provide our results in rather compact analytical formulae, which should be included in propagation-loss treatments aiming at reaching percent-level predictions around 1 GeV. 

ii) Additionally, we notice that all current popular CR codes (such as \texttt{GALPROP}~\footnote{https://galprop.stanford.edu}\cite{1998ApJ...493..694M}, \texttt{DRAGON}~\footnote{https://github.com/cosmicrays}, \texttt{USINE}~\footnote{https://lpsc.in2p3.fr/usine}\cite{Maurin:2018rmm},  and \texttt{PICARD}~\footnote{https://astro-staff.uibk.ac.at/\~{}kissmrbu/Picard.html}\cite{Kissmann:2014sia}) 
deal with $pp$ inelastic cross sections as a continuous energy loss channel, typically relying on the analytical formulation for the stopping power reported in~\cite{2015ApJ...802..114K}. 
We gauge the accuracy of this approximation, both for the inelastic and elastic channels, and find it insufficient for current precision, leading to $\sim 3\%$ errors. An iterative scheme for more correctly incorporating catastrophic losses is presented. 

iii) As an ancillary task, we also assess the level of error committed when neglecting sub-leading effects in ionization losses which, contrary to the above-mentioned processes, are electromagnetic in nature. These corrections to the leading effects (typically included by relying on the formulae presented in~\cite{Mannheim:1994sv}) are often small, at the 0.1\% level for protons, and may still be neglected. We note, however, that they are expected to be one order of magnitude bigger for iron, and neglecting them becomes then more questionable, especially for the precision expected in future measurements.

The structure of the article is the following.  In Sec.\ref{sec:Elosses} we describe the main analytical results of our article, pertaining the treatment of elastic $pp$ $E$-losses.
In Sec.~\ref{sec:Pfluxes} we describe the semi-analytical model used to gauge the effects on the CR energy fluxes, our approximations and iterative schemes.  Our main results are presented in Sec.\ref{sec:Results}, while Sec.~\ref{sec:conclusions} reports our conclusions and perspectives for future investigation. In the Appendix~\ref{appendix}, we  briefly introduce and display the subleading effects in ionization energy-losses for both protons and Fe nuclei.

\section{Proton-proton elastic cross section and cosmic ray energy losses}\label{sec:Elosses}
The essential ingredients required to leverage a two-body scattering process, $1+2\to3+4$, within the transport equation are the total and differential cross-sections associated with the process. In Sec.~\ref{xsec_el}, we report the  expressions for such cross sections for the $pp$-elastic process. In Sec.~\ref{StPow_el}, based on these results, we derive analytical relations for its \textit{Mass Stopping Power} and \textit{Inelasticity}, crucial derived quantities to incorporate into the transport equation, and compare the magnitude of the stopping power for elastic process to the other processes already accounted for. Unless stated otherwise, we use natural units in our symbolic equations. 

\subsection{The differential elastic $pp$ cross section.}\label{xsec_el}
As a preliminary step, we introduce the expressions of the Mandelstam variables, $s$, $u$ and $t$,  in terms of the incident particle's momentum $p_{\rm lab}$, the $i$-th body kinetic energy $K_i$ and the transferred energy $W \equiv E_1-E_3 = K_1-K_3$. In the Lab frame, where the body $2$ is at rest,  for the $p+p\to p+p$ reaction energy conservation implies 
\begin{equation}\label{eq:E_conserve_Lab}
    E_1+m_p=E_3+E_4\,,
\end{equation}
and one has additionally
\begin{equation}\label{eq:tMan}
    t = (p_1-p_3)^2=(p_2-p_4)^2=2m_p^2-2m_pE_4=-2m_p(E_4-m_p)=-2m_pK_4\,,
\end{equation}
where $E_i = K_i+m_p$ is the $i$-th proton's total energy. The  other variables of interest write $W=-t/2m_p$, $s = 2m_p\left(m_p+\sqrt{m_p^2+p_{\rm lab}^2}\right)=2m_p(2m_p+K),$ and $u = -2m_pK_3$.
Henceforth, we denote the kinetic energy of the projectile in the Lab frame simply as $K$.

The $pp$-elastic differential cross-section can be parameterized in terms of the Mandelstam variables as~\cite{Cugnon:1996kh}~\footnote{Here we are taking into account the symmetrised form, which is stated to be the correct one but not actually reported in~\cite{Cugnon:1996kh}. Also, we are neglecting the Coulomb term, which is negligible for measured CR energies $\gtrsim$0.1 GeV.} 

\begin{equation}\label{eq:dxsec_pp}
    \frac{\d\sigma}{\d t}=\mathcal{A}(s) \left[e^{B_{ pp}(s)\,t}+e^{B_{ pp}(s)\,u}\right]\,,
\end{equation}
where we  adopt the parameterization given in  eq. (8) of ref.~\cite{Cugnon:1996kh}, in units [${\rm (GeV/c)^{-2}}$]:
\begin{equation}\label{eq:B_ppel}
    B_{\rm pp}=\begin{cases}
    \frac{5.5p_{\rm lab}^8}{7.7+p_{\rm lab}^8}& (p_{\rm lab} <2{~\rm GeV/c^2})\\
    5.334+0.67(p_{\rm lab}-2) &(p_{\rm lab} >2{~\rm GeV/c^2})
\end{cases}\,.
\end{equation}

The normalization factor $\mathcal{A}(s)$ in eq.~\eqref{eq:dxsec_pp} can be determined by integrating the differential cross-section over $t$ to match the total elastic cross-section $\sigma_{\rm el}$, in turn parameterized in the following according to~\cite{Cugnon:1996kh}  

\begin{equation}
\sigma_{\rm el} = 
    \begin{cases}
     23.5 + 1000(p_{\rm lab}-0.7)^4 & p_{\rm lab}<0.8 \\
     \frac{1250}{p_{\rm lab}+50}-4(p_{\rm lab}-1.3)^2 & 0.8<p_{\rm lab}<2 \\
     \frac{77}{p_{\rm lab}+1.5} & p_{\rm lab}>2
\end{cases}\,.
\end{equation}

From the relation $t = -2m_pW$, we can write 
\begin{equation}
    \frac{\d\sigma}{\d W} = \frac{\d\sigma}{\d t}\frac{\d t}{\d W} =- 2m_p \frac{\d\sigma}{\d t}\,.
\end{equation}
Note that the eq.~\eqref{eq:dxsec_pp} exhibits symmetry under the exchange of particles $3\leftrightarrow$ 4. This symmetry is expected, reflecting the indistinguishability of the final state protons. Thus, to obtain the energy loss rate and the average kinetic energy fraction carried by the leading outgoing proton, we need to restrict the transferred energy range to [0,$K/2$]. Hence
\begin{equation}\label{normalxel_1}
\sigma_{\rm el}=\int_0^{K/2} \d W\,    \frac{\d\sigma}{\d W}=-2m_p{\mathcal{A}}\int_0^{K/2} \d W\,  \left\{e^{-2m_pB_{\rm pp}W}+e^{-2m_pB_{\rm pp}(K-W)}\right\}=-\frac{{\mathcal{A}}}{B_{pp}}\left(1-e^{-2m_pB_{pp}K}\right)\,,
\end{equation}
so that
\begin{equation}\label{dsdW_1}
    \frac{\d\sigma}{\d W} = 2\sigma_{\rm el}\,B_{pp}m_p \frac{e^{-2m_pB_{\rm pp}W}+e^{-2m_pB_{\rm pp}(K-W)}}{1-e^{-2m_pB_{pp}K}}\Theta\left(\frac{K}{2}-W\right)\,.
\end{equation}

\begin{figure}[t!]
\centering
\includegraphics[width=0.7\textwidth]{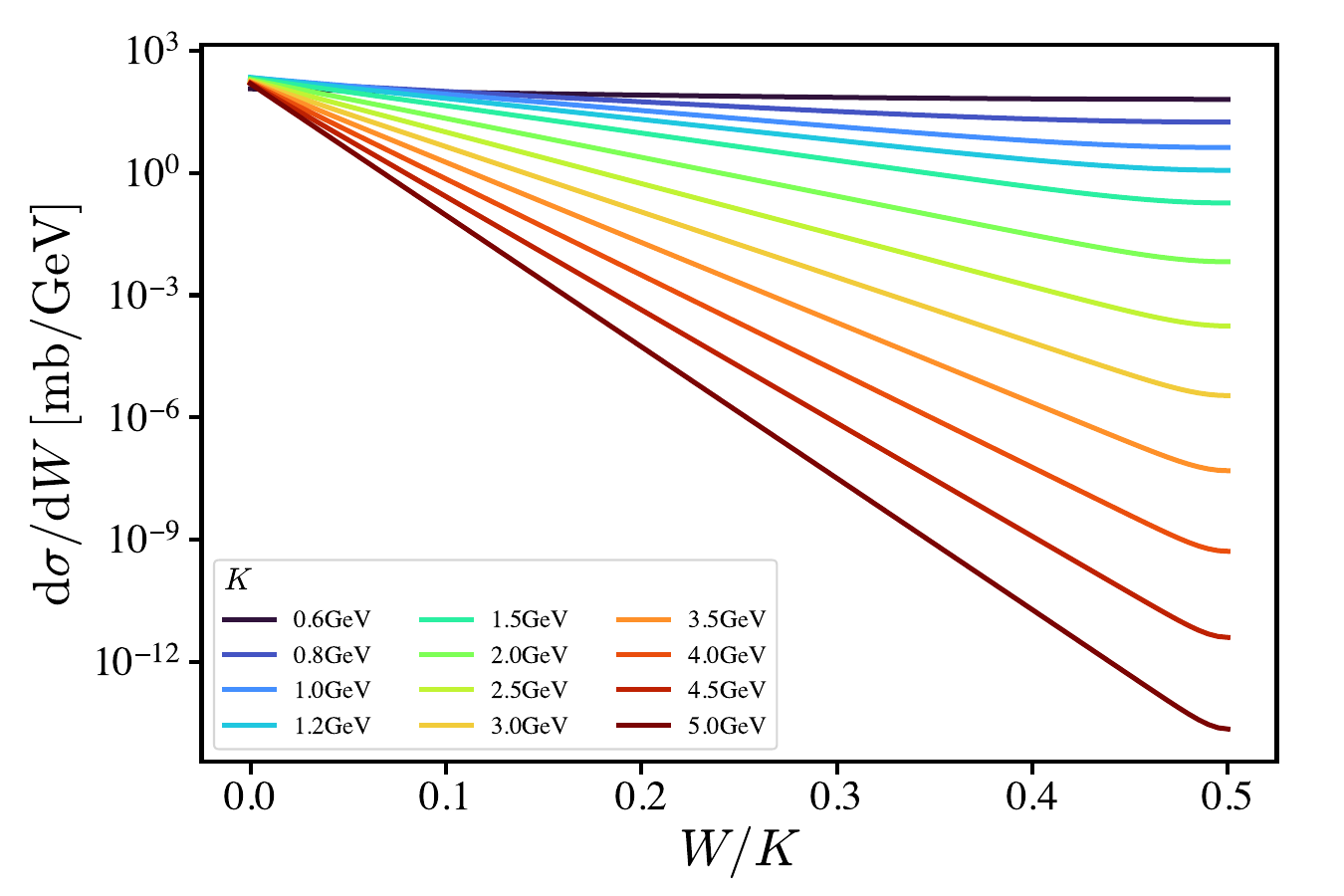}
\caption{ Differential energy-transfer cross-section, eq.~\eqref{dsdW_1}, as a function of $W/K$.}
\label{fig:1}
\end{figure}

This cross section, whose trend is illustrated in Fig.~\ref{fig:1}, is almost flat for non-relativistic proton energies, while more and more forward-peaked in the relativistic limit, without however being accompanied by a significant growth of its normalization: We expect thus that the most prominent effect for energy losses is obtained at low-energies, as we confirm in the next section. 

\subsection{Stopping power and inelasticity.}\label{StPow_el}
The primary input to assess the impact of any process onto CR energy losses
is the  {\it Stopping power}, which is defined for the process $i$ as 
\begin{equation}
 \left(- \frac{\d E}{\d x} \right)_i= n\int_0^{K_{\rm max}} \d W W\frac{\d\sigma_i}{\d W} \,,
\end{equation}
where $n$ is the interstellar medium target density. Since nuclear effects are beyond our interest here, in the following a pure hydrogen composition is assumed.

\begin{figure}[t!]
\centering
\subfloat[]{\includegraphics[width=0.5\textwidth]{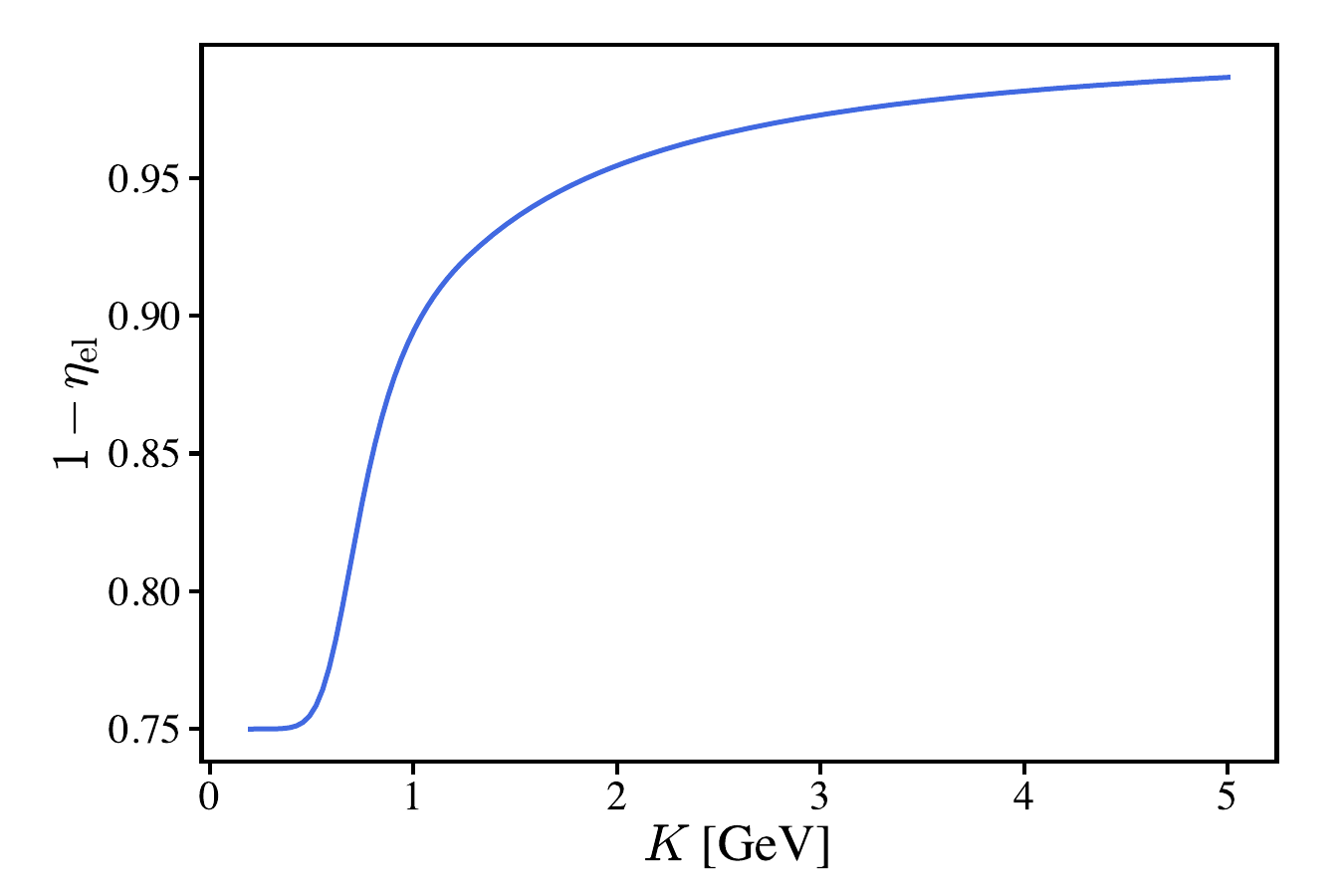}
\label{fig:2}}
\subfloat[]{\includegraphics[width=0.5\textwidth] {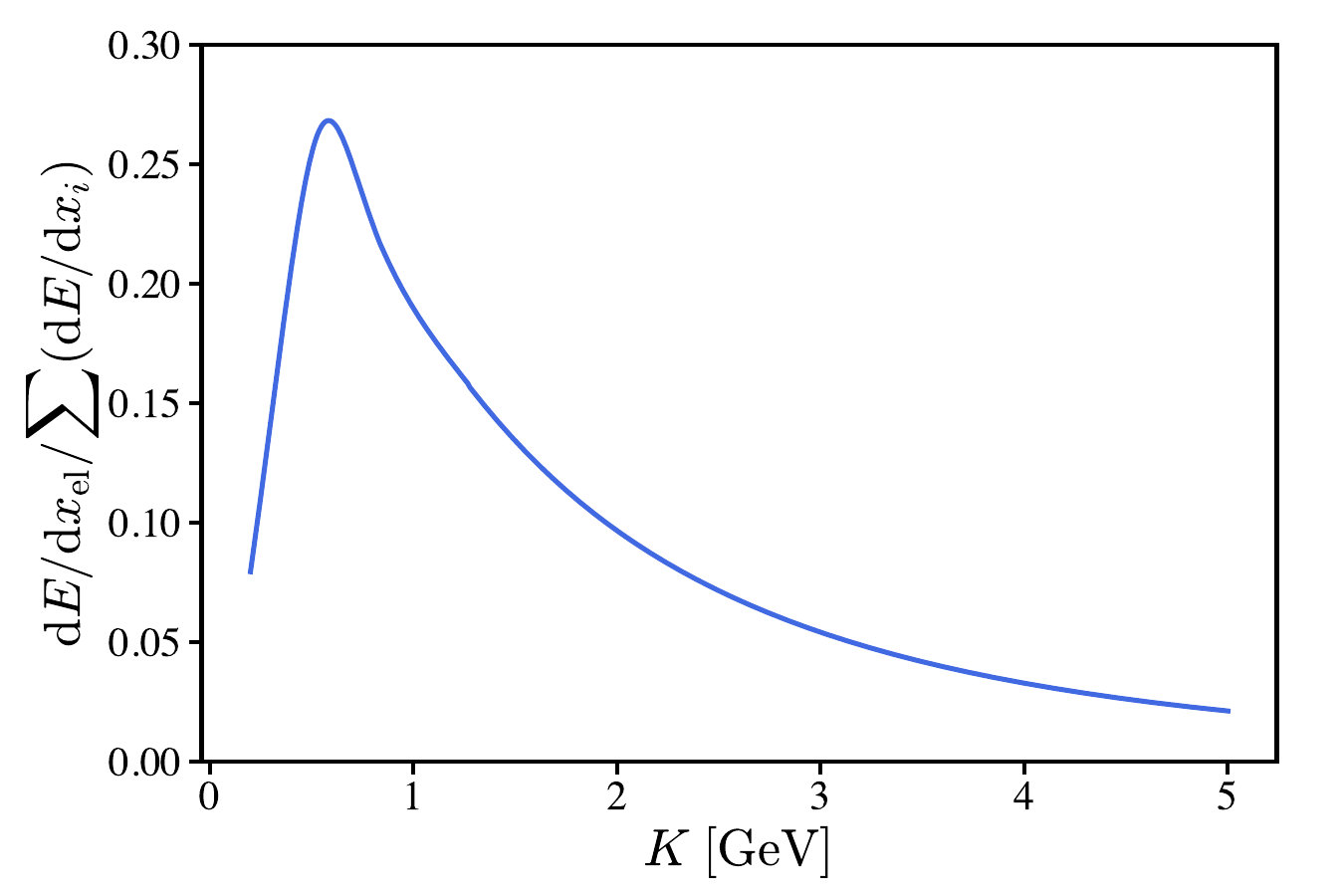}
\label{fig:3}}

\caption{(a) The average energy fraction retained in the collision, $1-\eta_{\rm el}$, from eq.~(\ref{av_E_tran}). (b) Relative weight of the elastic stopping power with respect to the total stopping power.
}

\end{figure}

Based on eq.~\eqref{dsdW_1}, we can compute:
\begin{equation}\label{stpowel1}
   \left(- \frac{\d E}{\d x} \right)_{\rm el} = -2m_p n\mathcal{A}\int_0^{K/2} \d W\, W\,  \left\{e^{-2m_pB_{\rm pp}W}+e^{-2m_pB_{\rm pp}(K-W)}\right\}\,=-\frac{n\mathcal{A}}{2B_{pp}^2m_p}e^{-2m_pB_{pp}K}\left(e^{m_pB_{pp}K}-1\right)^2\,,
\end{equation}
and using eq.~\eqref{normalxel_1}, we obtain
\begin{equation}\label{stpowel}
   \left(- \frac{\d E}{\d x} \right)_{\rm el} =\frac{n\,\sigma_{\rm el}}{2B_{pp}m_p}\tanh\left(\frac{B_{pp}m_pK}{2}\right)\,,
\end{equation}
as well as
\begin{equation}\label{av_E_tran}
\eta_{\rm el}\equiv\frac{\langle W\rangle_{\rm el}}{K}=\frac{\int_0^{K/2} \d W\,  W \frac{\d\sigma}{\d W}}{K\int_0^{K/2} \d W\,   \frac{\d\sigma}{\d W}}=\frac{\tanh\left(\frac{B_{pp}m_pK}{2}\right)}{2B_{pp}m_p\,K}\,,
\end{equation}
where we introduced the inelasticity $\eta$, such that $1-\eta$ is the average fraction of initial kinetic energy retained by the projectile after the collision. 
In  Fig.~\ref{fig:2}, we show this function vs. $K$ for the elastic channel, computed according to eq.~\eqref{av_E_tran}:  $\eta_{\rm el}$ is exceeding $\sim 10\%$ at $K\lesssim 1\,$GeV and is still of a few percent up to $\sim 5\,$GeV. In the following, we will focus on this energy range.

It is instructive to compare the above quantities with the analogue ones for competing processes. At low energies, the dominant CR proton loss channel is associated to ionization (and, to a less extent, Coulomb losses). 

For the ionization stopping power,  $(-\d E/\d x)_{\rm ion}$, we use the results from  the code \texttt{Crange}~\footnote{https://www.thedreamweaver.org/crange/index.html}, associated to the publication~\cite{Weaver:2002st}. See the Appendix~\ref{appendix} for further details.

Above its kinematical threshold at $K\simeq 290\,$MeV~\cite{Serpico:2023yag}, a key inelastic process affecting protons at low energies is (mostly single) pion production; its main effect is to 
drain on average $\kappa_\pi\simeq 17\%$ of the impinging proton kinetic energy into the produced pion, see~\cite{2000A&A...362..937A,Kelner:2006tc},
so that the resulting downscattered nucleon retains a kinetic energy fraction $\eta_{\rm in}\equiv 1-\kappa_\pi$.
The inelastic cross-section is obtained from Ref.~\cite{Cugnon:1996kh}, subtracting from the total cross section in eq. (1) the total elastic one in eq. (5). Within this approximation, one has 
\begin{equation}\label{stpowinel}
 \left(  -\frac{\d E}{\d x}\right)_{\rm in} =n\,\sigma_{\rm in}\kappa_\pi K\,.
\end{equation}

In Fig.~\ref{fig:3}, we see that the elastic interaction channel can exceed 20\% of the total energy-loss term, hence it makes sense to assess its impact on observable quantities. 

However, loss processes are not the dominant transport for protons, as non-collisional processes, notably diffusion, prevail. To assess the importance of the elastic interactions on the observable, i.e. the CR fluxes, we must introduce the transport equation and elucidate our approach to its solution. This is the goal of the following section.

\section{Assessing the relative effect on cosmic-ray proton fluxes}\label{sec:Pfluxes}

In a ``modified weighted slab'' propagation model (see e.g.~\cite{Jones:2000qd}, henceforth reported in the notation of~\cite{Aloisio:2013tda}), the proton flux at kinetic energy $K$, denoted by $I(K)$, can be obtained by solving the transport equation
\begin{equation}\label{eq:transport}
    \frac{I(K)}{X(K)}+\frac{\d}{\d K}\left[-\left(\frac{\d E}{\d X}\right)I(K)\right]= 2h_d\frac{A p^2q_{0}(p)}{\mu v}-\frac{\sigma_p I(K)}{m_p}+\sum_{\alpha'}\int {\rm d} K_{\alpha'}\frac{I_{\alpha'}(K_{\alpha'})}{m}\frac{{\rm d }\sigma_{\alpha'\to p}(K_{\alpha'},K)}{{\rm d} K_{\alpha'}}\,,
\end{equation}
where $p$ is the momentum ($p(K)= \sqrt{K^2+2K\,m_p}$), $\sigma_p$ is the total proton cross-section, ${\rm d }\sigma_{\alpha'\to p}/{\rm d} K_{\alpha'}$ is the differential cross-section for a nucleus $\alpha'$ of kinetic energy $K_{\alpha'}$ to yield a proton of kinetic energy $K$, $q_{0}(p)$ is the rate of injection per unit volume in the disc of the Galaxy, $v$ is the velocity of the nuclei type $\alpha$ (in units of $c=1$), $\mu = 2h_dn_dm$ is the surface gas density in the disc with $h_d$ being the half-thickness of the Galactic disc, $n_d$ is the gas number density in the disc, and 
\begin{equation}
    X(K) = \frac{\mu v }{2v_A}\left[1-\exp\left(-\frac{v_A}{D}H\right)\right]\label{eq:grammage}
\end{equation}
is the grammage experienced by protons. Here $v_A$ is advection velocity (in units of $c=1$), expected to be of the order of the Alfv\'en velocity, and $D$ is the diffusion coefficient for protons. For our illustrative purposes, we adopt $n_d=1~{\rm cm^{-3}}$
and $\mu = 2.4~{\rm mg/cm^2}$~\cite{Aloisio:2013tda}. 
The diffusion coefficient $D$ is a universal function (i.e. applies to all CR species) if expressed in terms of the rigidity $R$, i.e. momentum over charge, simply reducing to momentum for protons. In particular, 
we use the fits of the \texttt{BIG} and \texttt{SLIM} models discussed in~\cite{Genolini:2019ewc,Weinrich:2020ftb}, with the parameterisation 
\begin{equation}
    D(R) = \beta^\eta D_0\left\{1+\left(\frac{R}{R_l}\right)^\frac{\delta_l-\delta}{s_l}\right\}^{s_l}\left(\frac{R}{1\,{\rm GV}}\right)^\delta
    \left\{1+\left(\frac{R}{R_h}\right)^\frac{\delta-\delta_h}{s_h}\right\}^{-s_h}\,.
\end{equation}
We address the reader to~\cite{Weinrich:2020ftb} for the meaning and values of the parameters. 
Also, in this parametrization a halo size of $L = 5~{\rm kpc}$ is adopted.

The bracket in eq.~(\ref{eq:transport}) contains all approximately continuous energy-loss terms, expressed in terms of grammage, $ \d X=\rho\d x$, with $\rho=m_p\,n_d$ the ISM medium mass density. 

In the baseline model, we include
\begin{equation}\label{eq:SumdEdx}
   \left(\frac{\d E}{\d X}\right) = \left(\frac{\d E}{\d X}\right)_{\rm ad} + \left(\frac{\d E}{\d X}\right)_{\rm ion}
\equiv  -\mathcal{S}(K) \,,
\end{equation}
for the advection and ionization 
 loss terms, respectively. The advection stopping power is given by 
\begin{equation}\label{eq:dEdx_ad}
    \left(\frac{\d E}{\d X}\right)_{\rm ad} = -\frac{2v_A}{3\mu } \sqrt{K(K+m_{\rm p}c^2)}~. 
\end{equation}

To solve the propagation equation, we proceed as follows. First, we ignore all but continuous energy losses, which reduces the equation to
\begin{equation}\label{eq:transport0}
    \frac{I(K)}{X(K)}+\frac{\d}{\d K}\left[-\left(\frac{\d E}{\d X}\right)I(K)\right] = 2h_d\frac{A p^2q_{0}(p)}{\mu v}\,.
\end{equation}
Eq.~\eqref{eq:transport} can be re-written as
\begin{equation}\label{eq:transp_2}
    \Lambda_{1}(K)I+\Lambda_{2}(K) \frac{{\rm d} I}{{\rm d} K} = Q(K)~,
\end{equation}
where we assume
\begin{equation}
    Q(K) =\kappa \left(\frac{p(K)}{m_p}\right)^{-\gamma}\,,\label{red_eq}
\end{equation}
 
with $\gamma\simeq 2.2-2.4$, while the normalization factor $\kappa$ holds no significance for our purposes, as will be clear from the subsequent discussions. 
The coefficients in the eq. \eqref{eq:transp_2} can be explicitly written as
\begin{equation}\label{eq:lambda1}
    \Lambda_{1} = \frac{1}{X(K)}+\frac{\d\mathcal{S}}{\d K}
\end{equation}
and 
\begin{equation}\label{eq:lambda2}
    \Lambda_{2}(K) = \mathcal{S}(K)\,.
\end{equation}

Hence, the solution to eq.~(\ref{red_eq}), vanishing at $K\to \infty$, writes
\begin{equation}\label{eq:I_solution}
    I(K) = \int_K^\infty \d K' \frac{Q(K')}{\Lambda_{2}(K')}\exp\left[-\int_K^{K'} \d K''\frac{\Lambda_{1}(K'')}{\Lambda_{2}(K'')}\right]\,.
\end{equation}
In practice, since the integrand in the exponential is rather large, the following approximation is pretty accurate (typically at 0.1\% level or better):
\begin{equation}\label{eq:I_solutionappr2}
    I(K) \simeq\int_K^\infty \d K' \frac{Q(K')}{\Lambda_{2}(K')}\exp\left[-(K'-K)\frac{\Lambda_{1}((K+K')/2)}{\Lambda_{2}((K+K')/2))}\right]\,.
\end{equation}
For rough  expectations, a more radical approximation valid in the limit $K\Lambda_1\gg \Lambda_2$ is
\begin{equation}
    I(K)\simeq \frac{Q(K)}{\Lambda_1}\,.\label{eq:rough}
    \end{equation}

\begin{figure}[t!]
\centering
\includegraphics[width=0.7\textwidth] {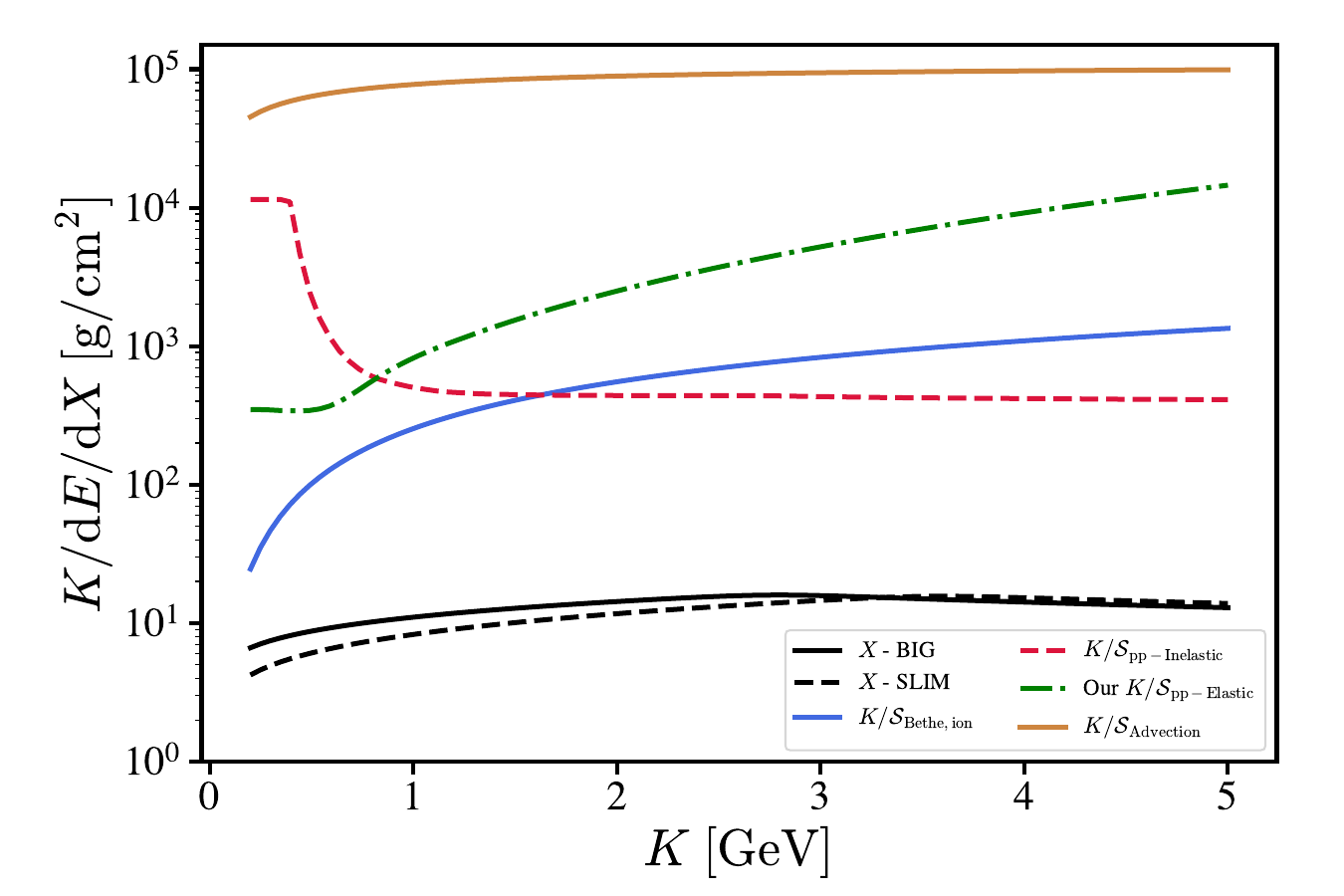}
\caption{Stopping range scale (in terms of grammage) associated to different energy loss channels, compared with the diffusive transport grammage for the \texttt{BIG} model.}
\label{fig:4}
\end{figure}

In Fig.~\ref{fig:4}, we illustrate the grammage of eq.~\eqref{eq:grammage} as well as the traditionally used proxies $K/{\cal S}_i$ for the loss terms entering the expression of $\Lambda_1$. Based on the rough expectation of eq.~\eqref{eq:rough}, we anticipate that losses contribute only at the ${\cal O}(10\%)$ level in shaping the spectrum. Hence, the effects on the CR flux observable should be one order of magnitude smaller than those gauged by merely comparing energy losses. 

Once the simplified solution of eq.~\eqref{eq:I_solutionappr2}, henceforth denoted with $I_0$, has been obtained, an iterative approach is adopted to account for catastrophic energy-loss terms, neglected till now. 

Within the approximation described in Sec.~\ref{StPow_el}, the loss and gain terms associated to inelastic, pion production channel can be described by changing the RHS of eq.~\eqref{eq:transp_2} into

\begin{equation}\label{eq:pert-iter}
    Q_i(K)= Q(K)-\left[\frac{\sigma_{\rm in}(K)}{m_p}I_{i-1}(K)-\frac{\sigma_{\rm in}\left(\frac{K}{\eta_{\rm in}}\right)}{\eta_{\rm in}m_p}I_{i-1}\left(\frac{K}{\eta_{\rm in}}\right)\right]\,,
\end{equation}
at iteration stage $i=1,2,3,\ldots$ Usually three iterations are sufficient to obtain convergence at the $\sim0.1\%$ level. 

\begin{figure}[t!]
\centering
\includegraphics[width=0.7\textwidth] {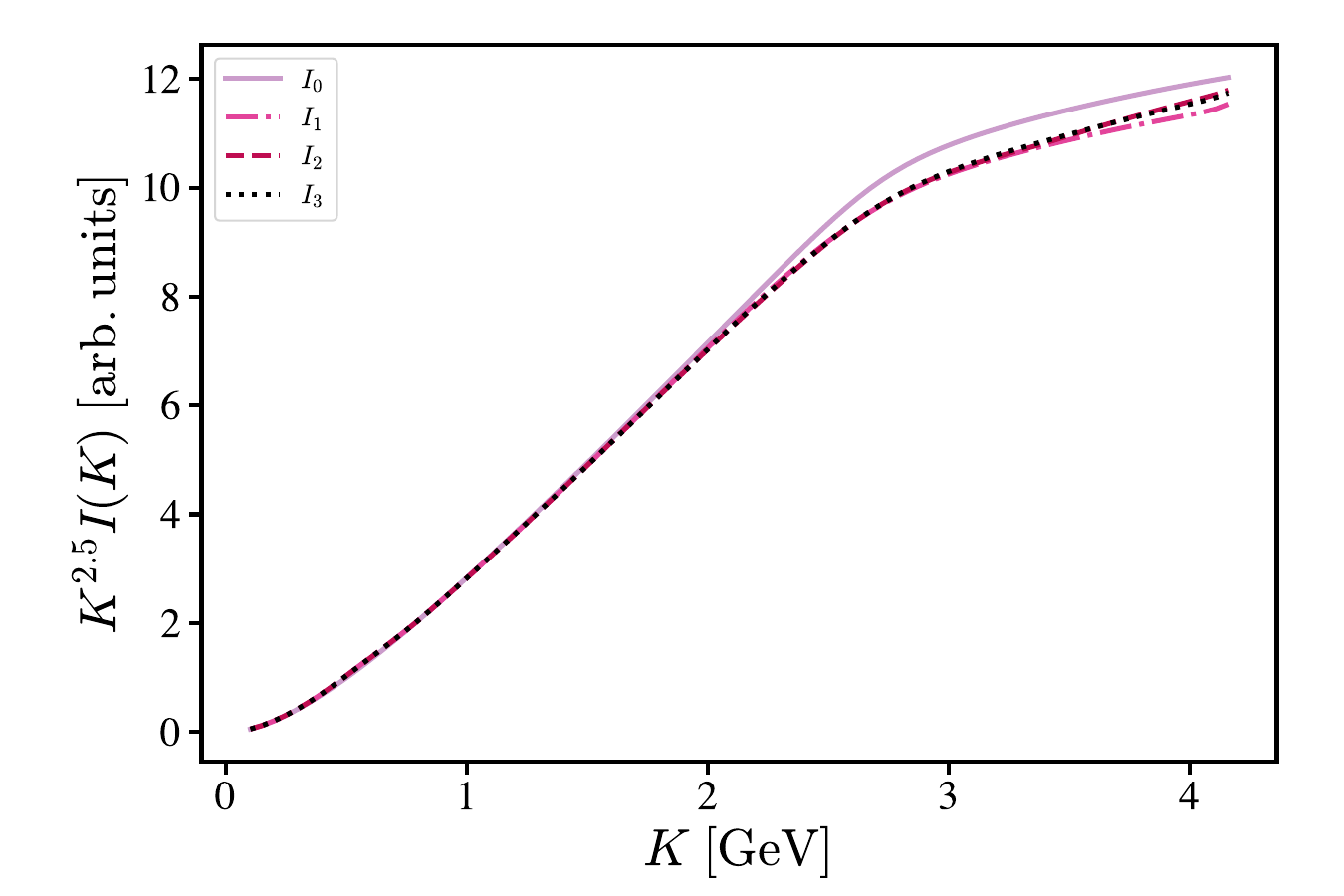}

\caption{Illustration of the iterative solution using eqs. (\ref{eq:pert-iterALL}) and (\ref{eq:I_solutionappr2}). A rapid convergence is manifest. See the text for the details.}
\label{fig:5}
\end{figure}

In order to account for the elastic cross section process, we follow a similar approach, just now  
\begin{equation}\label{eq:pert-iterALL}
    Q_i(K)= Q(K)-\left[\frac{\sigma_{\rm in}(K)}{m_p}I_{i-1}(K)-\frac{\sigma_{\rm in}\left(\frac{K}{\eta_{\rm in}}\right)}{\eta_{\rm in}m_p}I_{i-1}\left(\frac{K}{\eta_{\rm in}}\right)\right]-\left[\frac{\sigma_{\rm el}(K)}{m_p}I_{i-1}(K)-\frac{\sigma_{\rm el}\left(\frac{K}{\eta_{\rm el}}\right)}{\eta_{\rm el}m_p}I_{i-1}\left(\frac{K}{\eta_{\rm el}}\right)\right]\,.
\end{equation}
The rapid convergence of this method is illustrated in Fig.~\ref{fig:5}, for the case  $\gamma=2.2$ and the \texttt{BIG} model.

In this treatment, the main approximation consists in neglecting {\it secondary} protons coming from spallations of heavier CR nuclei, such as He, C, O\ldots Fe, allowing us to reduce the problem to the integration of a single differential equation, instead of a coupled system. Since we are only interested in the relative effect of including the elastic process, and both primaries and secondary protons undergo losses, accounting for secondaries from nuclear spallations is largely degenerate with the injection spectrum (or index) effect, which we briefly discuss in the next section.

\section{Results}\label{sec:Results}

\begin{figure}[t!]
\centering
\includegraphics[width=0.7\textwidth]{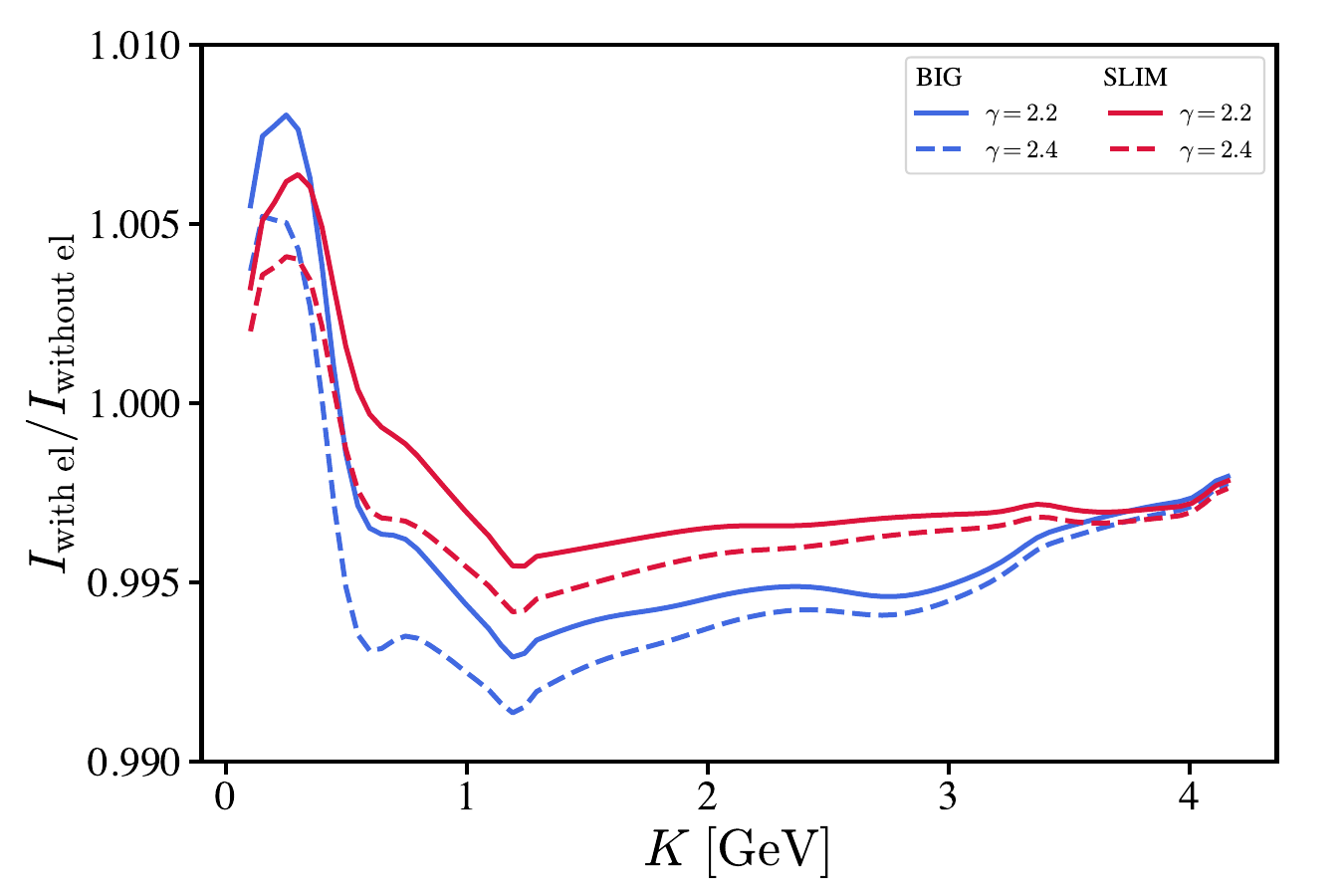}

\caption{Relative effect of including vs. neglecting the $pp$-elastic process, for the \texttt{BIG} (blue) and \texttt{SLIM} (red) models, and for two injection spectral indices, 2.2 (solid) and 2.4 (dashed).
} 
\label{fig:6}
\end{figure}

In Fig.~\ref{fig:6}, we illustrate the relative effect of the 
inclusion of the elastic loss process with respect to neglecting it, for two propagation models and two injection spectra. 
The shape of the curves is easy to explain: The process takes protons at higher energies and re-injects them at lower ones, hence the peak-dip structure.  Its magnitude is of the order of 0.8\% slightly below or slightly above 1 GeV, and should therefore be included whenever the ambition is stated to control the theoretical error to below such a level. The effect is similar in the two propagation models considered: A bit smaller in the \texttt{SLIM} model because the relative weight of the diffusion coefficient is more pronounced at low energies. Also note that steeper power-laws reduce the relative re-population effect and enhance the depletion one, as intuitively expected. 
The effect of including secondary protons should be similar, as discussed in the previous section. 

\begin{figure}[t!]
\centering
\includegraphics[width=0.7\textwidth]{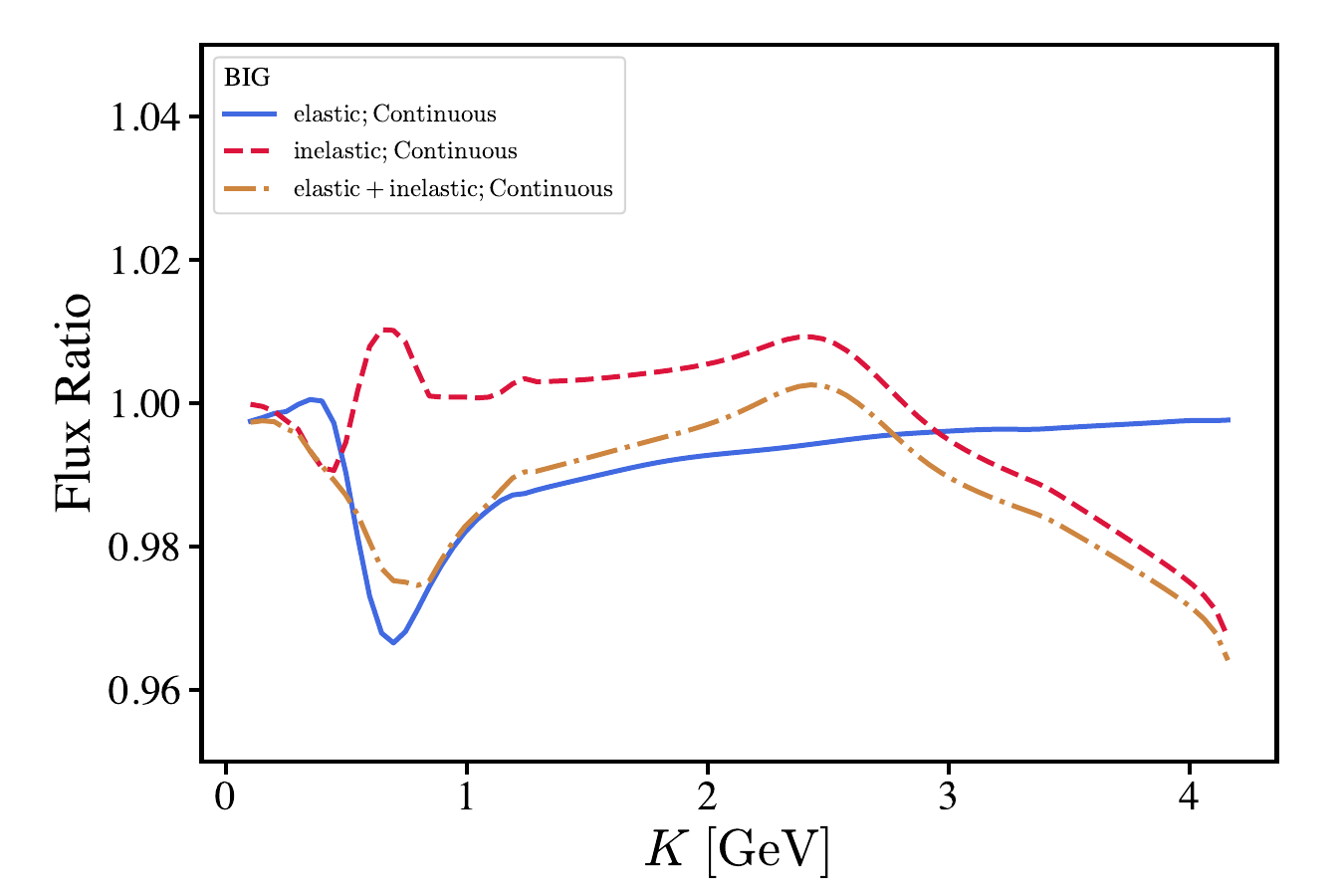}

\label{fig:CattoCont2}

\caption{CR proton flux ratio between the solution obtained with the continuous approximation for the elastic channel (blue, solid), the inelastic channel (dashed, red) or both (dot-dashed) over the solution obtained with the catastrophic treatment for both channels.
} 
\label{fig:7}
\end{figure}

In the limit where $\eta\to 0$, we expect that the ``catastrophic'' energy losses described via the brackets in eq.~\eqref{eq:pert-iterALL} admit a simpler description in terms of  a {\it continuous} energy loss mechanism. We can easily switch to such a treatment for the elastic channel, for instance, by setting the elastic channel bracket in eq.~\eqref{eq:pert-iterALL} to zero while  modifying 
\begin{equation}\label{eq:SumdEdxEL}
  \mathcal{S}(K)\to \mathcal{S}(K)=- \left(\frac{\d E}{\d X}\right)_{\rm ad}- \left(\frac{\d E}{\d X}\right)_{\rm ion}-\left(\frac{\d E}{\d X}\right)_{\rm el}\,.
\end{equation}
This change applies to the functions $\Lambda_1$ and $\Lambda_2$, as a consequence of the new function $\mathcal{S}(K)$ entering them. Of course, we can apply a similar treatment to the inelastic channel (together with or alternatively to the elastic one). In fact, current studies via popular CR propagation codes do adopt this approximation for pion production losses. 
In Fig.~\ref{fig:7} we see the effect of this continuous approximation. We see that it is definitely inappropriate for current accuracy goals, with errors reaching 3\%. It is easy to explain the different energy behaviour for the elastic and inelastic cases: For the elastic case, the biggest effect is when the elastic process matters the most, basically, just below 1 GeV, reflecting the drop of the inelasticity well above that value. For the pion production, the inelasticity is constant (within our approximation): The impact is higher at higher energies just because the pion-production process becomes the dominant loss channel there   (see Fig.~\ref{fig:4}).

\section{Discussion and conclusions}\label{sec:conclusions}
In the ongoing effort towards refining theoretical predictions for CR studies, we have focused in this article on the energy loss processes affecting protons, particularly relevant at low energies. Our main driver has been to raise awareness of the insufficient accuracy with which known physics is currently treated. 
We have scrutinised the currently used approximation in the field, notably the Bethe limit of the ionization energy loss, the neglecting of the elastic energy losses, and the continuous approximation to treat what are more exactly described as catastrophic losses.
While we confirm that, at least for protons, the former is sufficient, the latter two effects (and definitely the latter one, reaching 3\%) should be accounted for precision studies:
Note that these are all leading to energy-dependent effects, not reducible to a normalization uncertainty.
We have also provided the reader with compact analytical formulae for the quantities of interest for the elastic collision process, and shown that an iterative approach can successfully be used to tackle the last issue. 

This is but a step in a more extended effort, of course. A natural follow-up would be to incorporate these inputs in existing codes, and performing data analyses with/without the processes included to directly assess the systematics on astrophysical parameters of interest. This would also naturally account for secondary proton sources. Extending the treatment of these collisional losses to nuclei is another natural direction.

\begin{figure}[t!]
\centering
\includegraphics[width=0.7\textwidth]{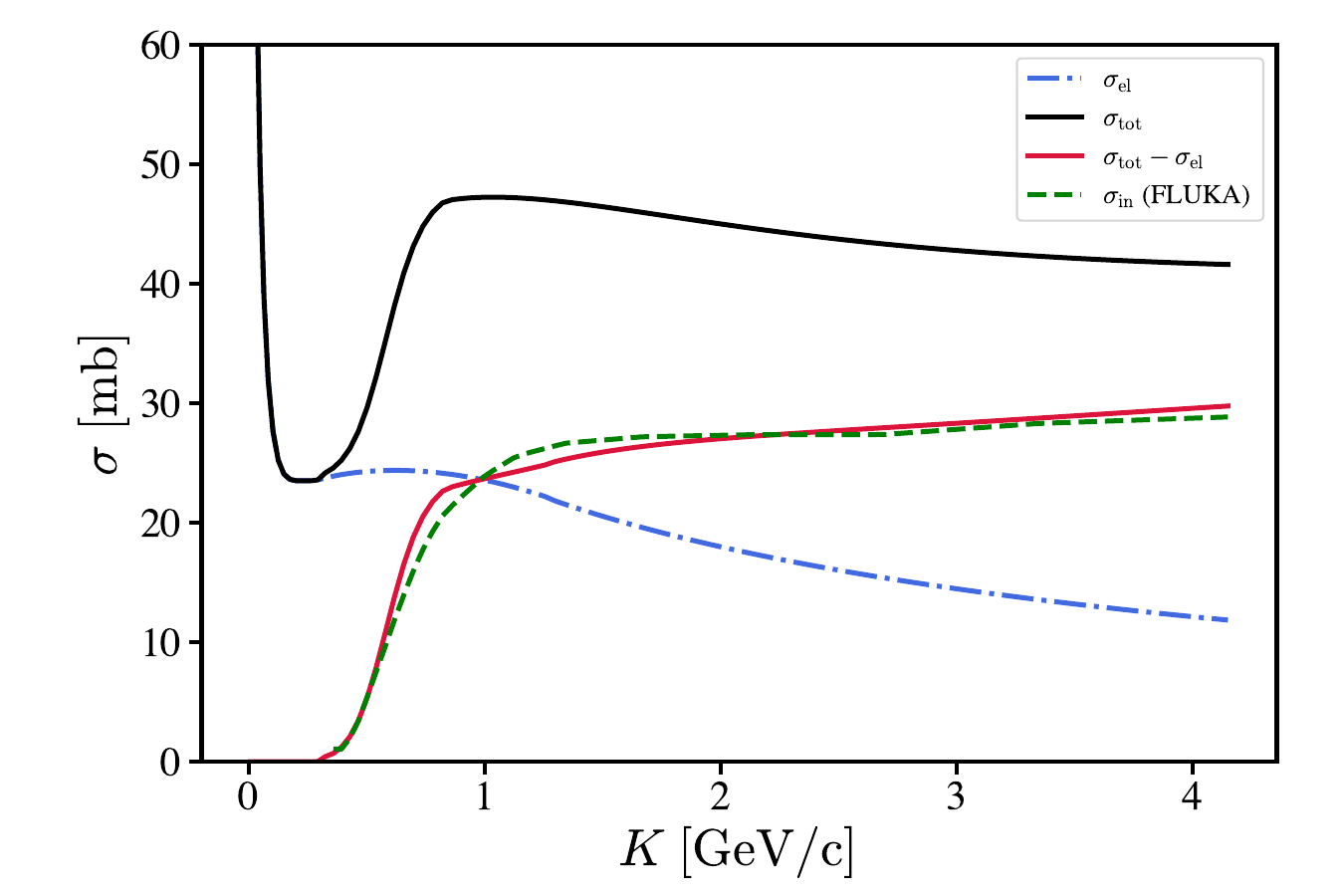}
\caption{total (black) and $pp$-elastic (blue) cross-section from \cite{Cugnon:1996kh} and their difference, i.e. the $pp$-inelastic (red). Comparison with the $pp$-inelastic from \texttt{FLUKA} as reported in~\cite{Mazziotta:2015uba} (green dashed).\label{fig:C1} }
\end{figure}

\begin{figure}[t!]
\centering
\includegraphics[width=0.7\textwidth]{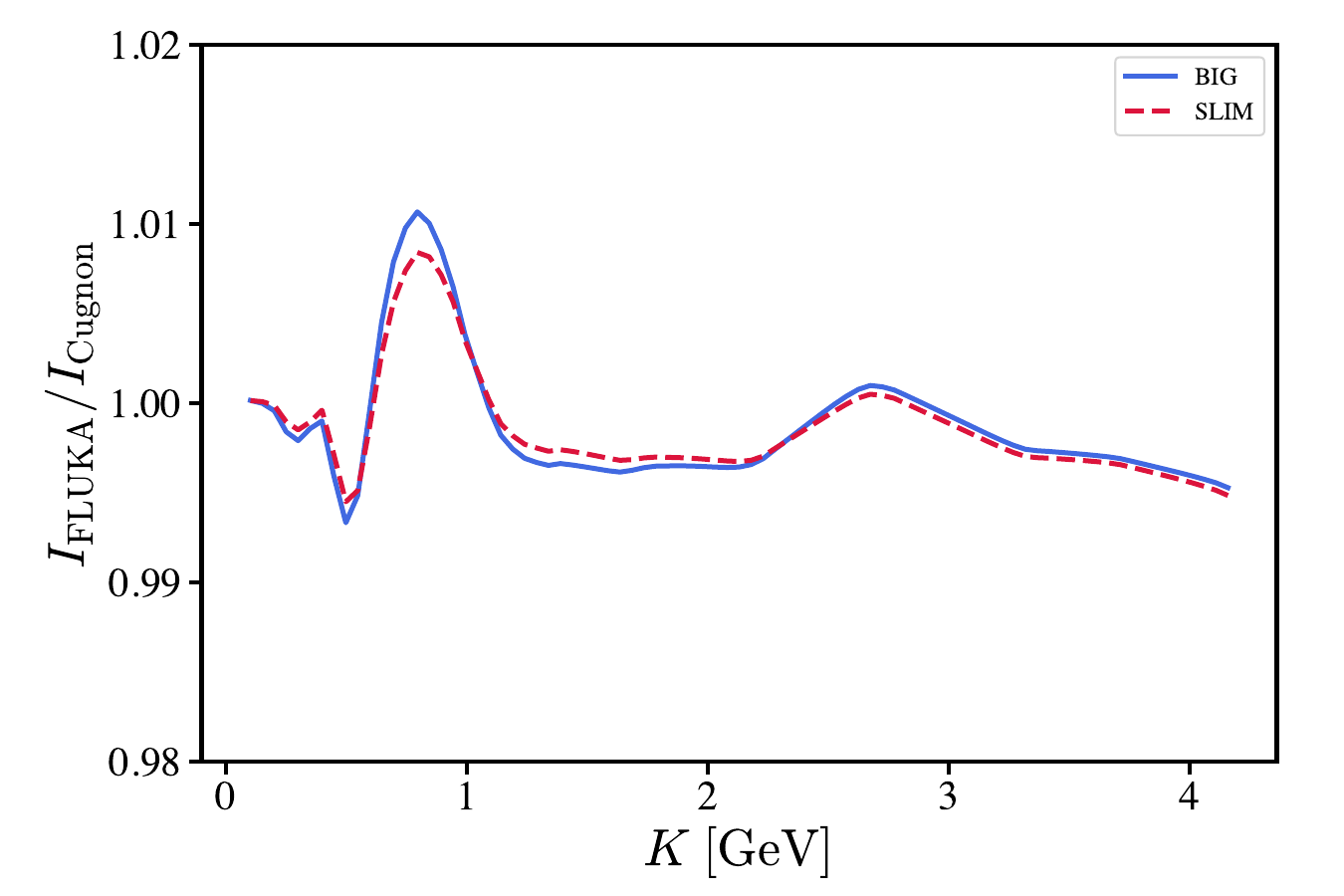}
\caption{Ratio of CR proton fluxes obtained by using the \texttt{FLUKA} $pp$-inelastic cross-section used  in~\cite{Mazziotta:2015uba} to the one from~\cite{Cugnon:1996kh} which is used in this article, i.e. the green-dashed curve with respect to the solid red curve in Fig.~\ref{fig:C1}. \label{fig:C2}}
\end{figure}

A further avenue consists in improving over existing parameterizations and fits of cross-section data, notably if new laboratory data should be available, and assess the errors affecting any parameterization used. 
As a preliminary step, we have compared the cross-section for the inelastic channel used here to the \texttt{FLUKA}~\footnote{\texttt{http://www.fluka.org}}-based cross section~\cite{Battistoni:2015epi} used e.g. in~\cite{Mazziotta:2015uba}, reported in Fig.~\ref{fig:C1}.  The corresponding effect on fluxes is displayed in Fig.~\ref{fig:C2}. We see that, despite the differences in fluxes being typically at around the 0.1\% level, differences up to 1\% can arise just above the threshold, and should be further examined. In particular, a trade-off may be at play between fits applicable over wide energy ranges and locally optimal fits. For instance, the single-pion production and the assumption of a fixed energy transfer $\kappa_\pi$ into the produced pion become inadequate at high-energies. Additionally, since these processes are stochastic, one may explore the role of fluctuations in collision via a  Monte Carlo study.

There are also other processes that we have not included. In particular, we have neglected the spallations that protons induce on interstellar medium nuclei. Accounting for these would not change qualitatively the above approach nor quantitatively the assessment for the effects we have considered. Our expectations is that, despite the fact that the cross sections for the process $p$-${}^4$He, say, reach $\sim 300\,$mb, only a very small fraction of the projectile proton ends up in the scattered products, i.e. the inelasticity is very small (see e.g. Figs. 4 and 6 in~\cite{Mazziotta:2015uba}). Yet, while these processes are taken into account as secondary nuclei sources, it is fair to say that neither the energy-loss effect on the projectile nor the  proton source  from target spallation are taken into account. These are all issues that we plan to tackle in future works. 

\begin{acknowledgments}

We thank Pasquale Blasi for his valuable feedback on the manuscript. A.F.~E. acknowledges support by the Fundação Carlos Chagas Filho de Amparo à Pesquisa do Estado do Rio de Janeiro (FAPERJ) scholarship No. 201.293/2023, the Conselho Nacional de Desenvolvimento Científico e Tecnológico (CNPq) scholarship No. 140315/2022-5 and by the Coordenação de Aperfeiçoamento de Pessoal de Nível Superior (CAPES)/Programa de Excelência Acadêmica (PROEX) scholarship No. 88887.617120/2021-00. P.D.S. has been partially supported by the CAPES-PRINT program No. 41/2017. A. E. thanks partial financial support by the Brazilian funding agency CNPq (grant 407149/2021).
\end{acknowledgments}

\appendix
\section{Sub-leading effects in the ionization losses}\label{appendix}

In each electromagnetic interaction with an electron of the medium, a proton loses a tiny fraction of its energy, hence a continuous energy-loss approximation is justified. The stopping power for ionization for a CR nucleus of charge $Ze$
takes the well-known form
\begin{equation}
   \left(- \frac{\d E}{\d x} \right)_{\rm ion} = \frac{4\pi n_e Z^2\alpha^2}{ m_e \beta^2}L\,,\:\:\:L\simeq L_{\rm Bethe}\left [\ln\left( \frac{2m_e\beta^2\gamma^2}{\cal I}\right)-\beta^2\right]\label{eq:BB}\,,
\end{equation}
where the expression given for $L$ corresponds to the result obtained by Bethe (${\cal I}$ being the effective ionization potential of the target). Although current treatments in CR astrophysics limit themselves to this approximation, see e.g. appendix C.10.4 in~\cite{Evoli:2016xgn}, more refined calculations for the dimensionless function $L$ are available. Some of these corrections are discussed in~\cite{Mannheim:1994sv}, but typically neglected without quantitative assessment of their magnitude. 
Here, we tackle this task, following the treatment given in~\cite{Weaver:2002st},  associated to the code \texttt{CRange}~\footnote{https://www.thedreamweaver.org/crange/recalc.html} which we use for numerical evaluation accounting for the following additional effects (with two-letters code to label them in parentheses):
\begin{itemize}
    \item Density effect (New Delta, ND)
    \item Lindhard-Sørensen correction (LS)
    \item Radiative correction (RA)
    \item Finite Nuclear Size (NS)
    \item Barkas Effect (BA)
    \item Shell Effect (SH)
    \item Leung Effect (LE)
    \item Modern Electron Capture Effect (EC)
    \item Kinematic Correction (KI)
    \item Pair Production (PA)
    \item Bremsstrahlung (BR)
\end{itemize}
We refrain here from a theoretical description of these effects, which goes well beyond the goals of our article. A modern exposition can be found for instance in~\cite{2022PhRvA.106c2809S}.  However, as a warning to the reader, we point out that, while \texttt{CRange} does nominally include a density effect correction  for interstellar medium and Galactic Halo conditions, its implementation is flawed and should not be used~\footnote{We would like to thank Cypris Plantier for pointing out some anomalous output associated to these choices, that triggered our investigation.}. Following the results reported in~\cite{Jackson:1998nia}, Sec. 13.3, 
we checked  that contrarily to what an uncritical  use of the code would indicate, this effect is completely negligible for CR propagation conditions (i.e. in a very rarefied medium) and can be safely neglected. 

\begin{figure}[t!]
\centering
\includegraphics[width=0.7\textwidth]{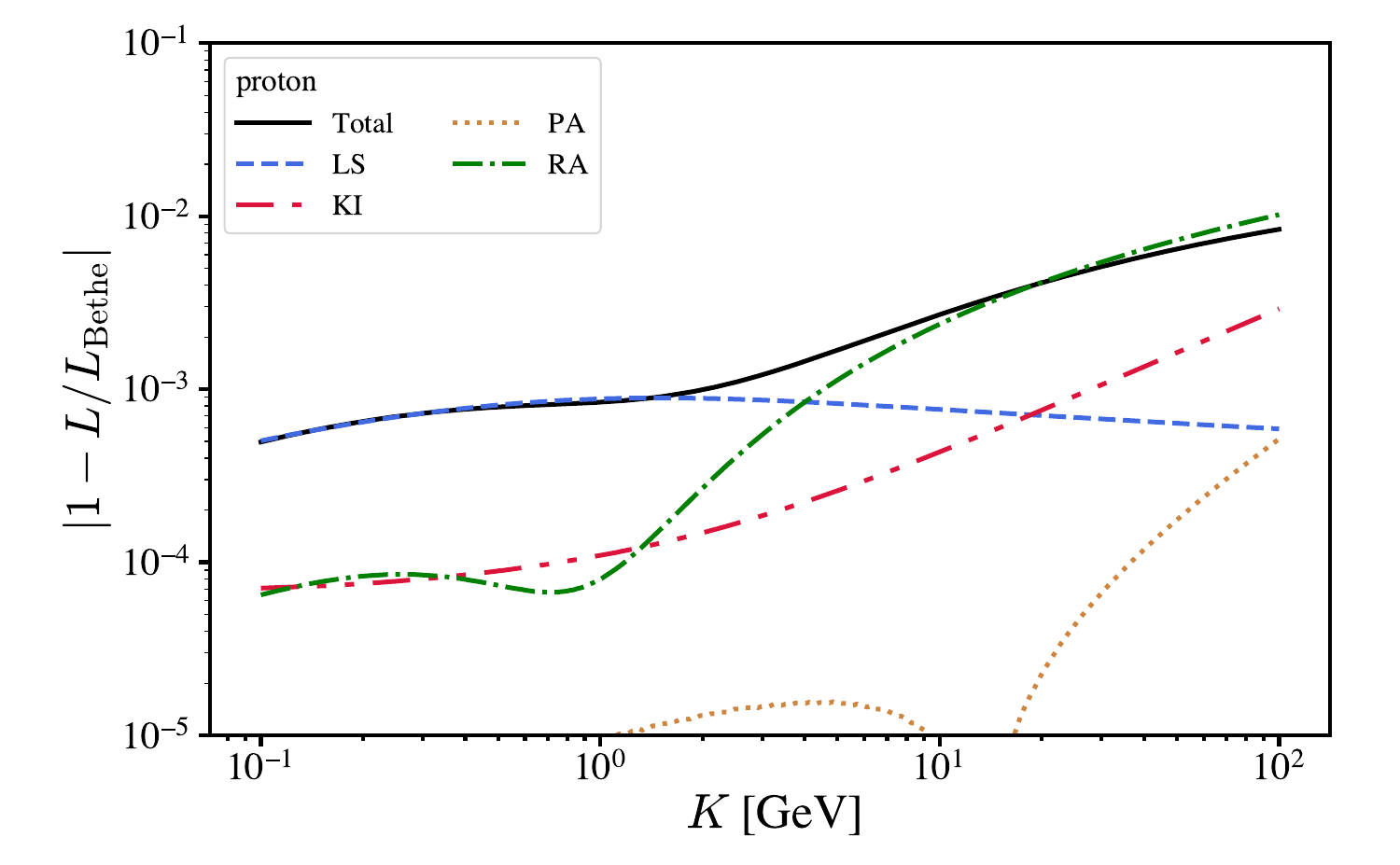}
\caption{Fractional corrections to the Bethe ionization stopping power for CR protons. See text for the labels of the different effects.\label{fig:A_1}}
\end{figure}

In Fig.~\ref{fig:A_1} we show the relative correction to the eq.~\eqref{eq:BB} due to different effects. 
The only one worth mentioning is the Lindhard-Sørensen correction~\cite{Lindhard:1996zz}, since the radiative correction dominating above 4 GeV (but still sub-percent!) intervenes at a point where ionization losses are already sub-leading (see Fig.~\ref{fig:4}). The LS-correction accounts for the solution of the relativistic Dirac-Coulomb equation, as opposed to the non-relativistic quantum-mechanical treatment of Bethe's results. Anyway, since the corrections are at the level of 0.1\%, we conclude that they can be safely ignored at the present accuracy goal.

\begin{figure}[t!]
\centering
\includegraphics[width=0.7\textwidth]{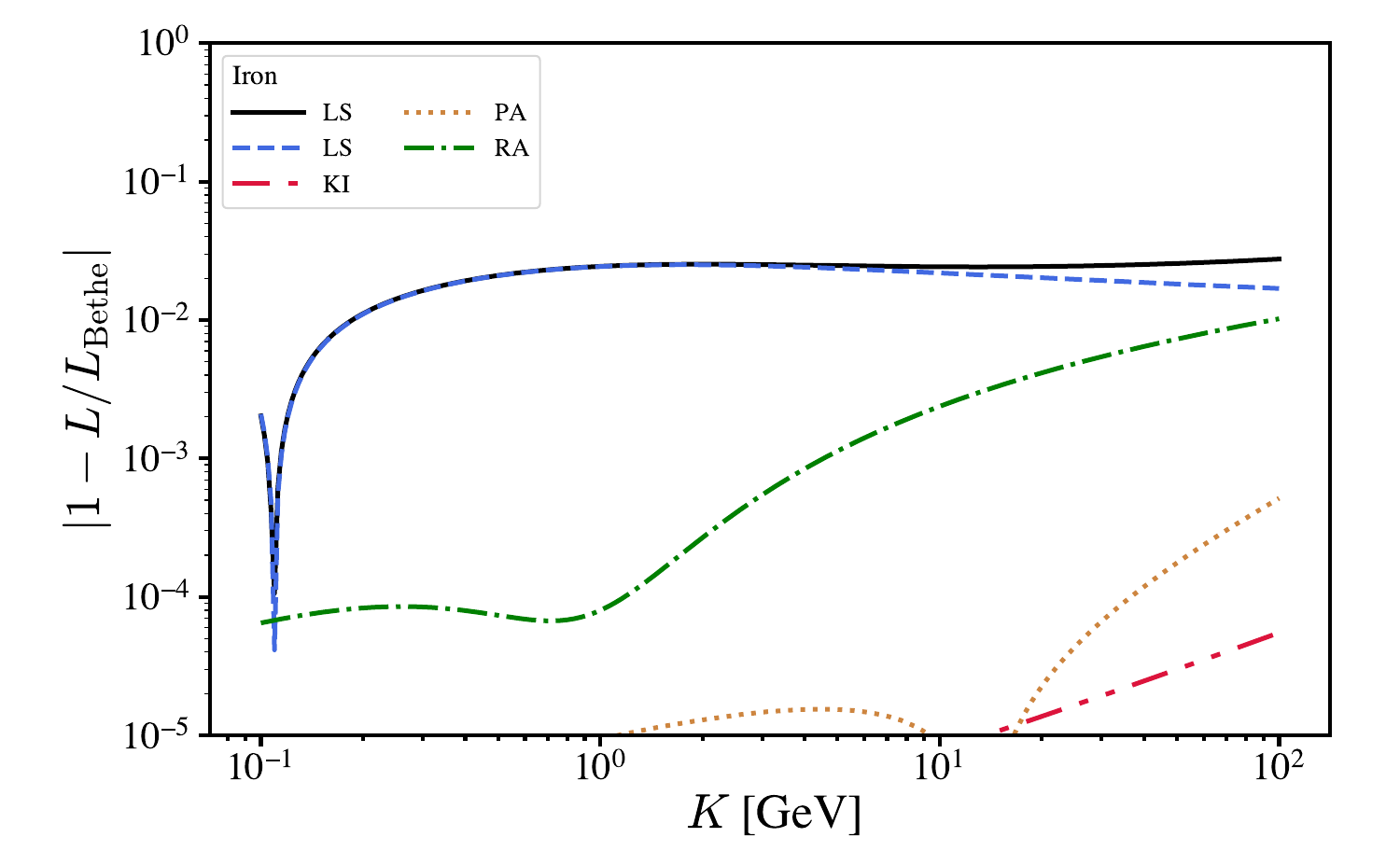}
\caption{Fractional corrections to the Bethe ionization stopping power for CR Fe.}
\label{fig:A_2}
\end{figure}

Although in this article we have focused  on CR protons, the electromagnetic corrections dealt with in this appendix are common to all nuclei, with a  known parametric dependence on the charge, mass, and size of the nucleus. We have thus explored how the above conclusions are altered in the case of Fe, which is the heaviest nucleus with a sizable abundance among CRs. As far as energy losses are concerned, we present our results in Fig.~\ref{fig:A_2}: we see that the LS correction reaches about 2\%.  

\begin{figure}[t!]
\centering
\includegraphics[width=0.7\textwidth]{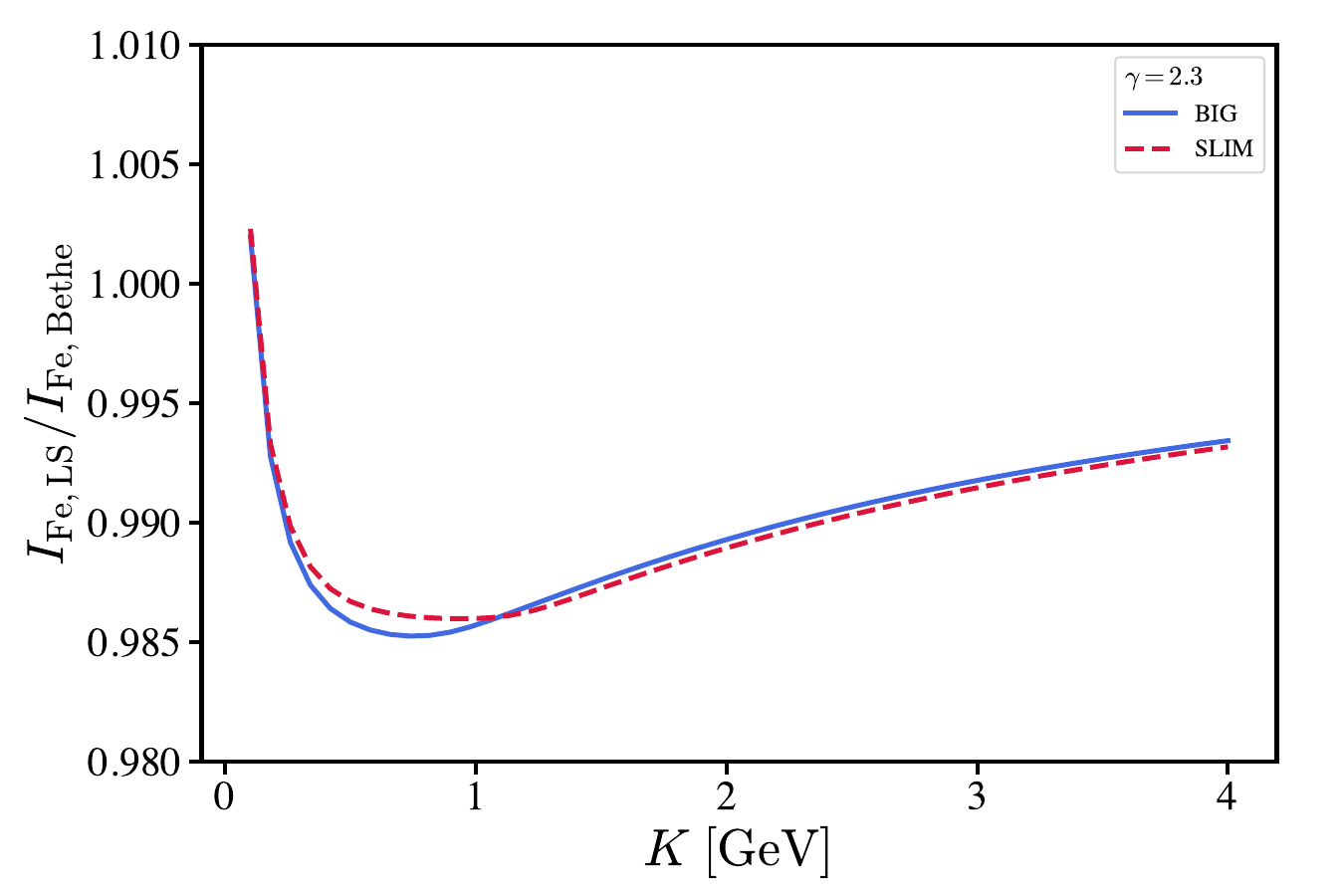}
\caption{ Iron flux ratio: The effect of the LS correction to the Bethe ionization expression in the \texttt{BIG} and \texttt{SLIM} model with $\gamma=2.3$.}
\label{fig:A_3}
\end{figure}

To gauge the impact on fluxes, we also solve the propagation equation for Fe. Technically, in this case we do not adopt an  iterative approach to solve the propagation equation, since the catastrophic loss channel is dominated by spallations onto the interstellar medium protons, but these are not associated to a sizable injection term counterpart, since iron is the heaviest nucleus with appreciable CR flux. Hence, we simply integrate eq.~\eqref{eq:I_solutionappr2} taking  
\begin{equation}\label{eq:iron}
     \Lambda_{1,{\rm Fe}} = \frac{1}{X_{\rm Fe}(K)}+\frac{\d\mathcal{S}_{\rm Fe}}{\d K}+\frac{\sigma_{\rm in}^{\rm Fe}(K)}{m_{\rm Fe}}\,,
\end{equation}
where Fe cross-section with $H$ target ($\sigma^{\rm Fe}_{\rm in}$) is taken from Table II of~\citep{Webber:1990ka}.
The results are summarised in Fig.~\ref{fig:A_3}. We see that the LS correction reaches  1.5\% just below 1 GeV, larger for instance than the statistical errors (albeit not systematic ones) on Fe data taken by AMS-02~\cite{AMS:2021lxc}.

\bibliography{refs}
\end{document}